\documentclass{article}
\usepackage{amsmath}
\usepackage{cite}
\usepackage{graphicx}
\usepackage{dcolumn}

\begin{document}

\date{}
\title{On the application of the Rayleigh-Ritz method to a projected Hamiltonian}
\author{Francisco M. Fern\'{a}ndez\thanks{%
fernande@quimica.unlp.edu.ar} \\
%EndAName
INIFTA, DQT, Sucursal 4, C. C. 16, \\
1900 La Plata, Argentina}
\maketitle

\begin{abstract}
We apply the well known Rayleigh-Ritz method (RRM) to the projection of a
Hamiltonian operator chosen recently for the extension of the Rayleigh-Ritz
variational principle to ensemble states. By means of a toy model we show
that the RRM eigenvalues approach to those of the projected Hamiltonian from
below in most cases. We also discuss the effect of an energy shift and the
projection of the identity operator.
\end{abstract}

\section{Introduction}

\label{sec:intro}

The Rayleigh-Ritz method (RRM) is one of the most widely used approaches for
the study of the electronic structure of atoms and molecules\cite{P68,SO96}.
One of its main advantages is that the RRM eigenvalues converge from above
towards the exact energies of the physical system\cite{M33} (see also\cite
{F22} and references therein).

In a recent paper, Ding et al\cite{DHS24} proposed an extension of the
Rayleigh-Ritz variational principle to ensemble states. In their
introduction to the principle for pure states they resorted to a most
curious Hamiltonian operator on a $D$-dimensional Hilbert space. Although
such an operator is quite unrealistic for the treatment of actual physical
problems, it seems to be worth further investigation. In section~\ref
{sec:unrealistic} we outline the formulation of the Rayleigh-Ritz
variational principle as given by Ding et al. In section~\ref{sec:RRM} we
discuss the application of the RRM to the projected Hamiltonian introduced
by those authors. Finally, in section~\ref{sec:conclusions} se summarize the
main results and draw conclusions.

\section{The unrealistic problem}

\label{sec:unrealistic}

In order to avoid misconceptions we resort to the exact expression of Ding
et al: ``As far as the computation of ground states is concerned, the
effectiveness of numerous numerical methods rests on the underlying
Rayleigh-Ritz (RR) variational principle: For any Hamiltonian
\begin{equation}
\hat{H}=\sum_{k=0}^{D-1}E_{k}\left| \psi _{k}\right\rangle \left\langle \psi
_{k}\right|  \label{eq:H_DHS}
\end{equation}
on a $D$-dimensional Hilbert space $\mathcal{H}$ the ground state energy $%
E_{0}$ can be obtained by variational means according to
\begin{equation}
E_{0}\leq \left\langle \psi \right| \hat{H}\left| \psi \right\rangle
,\;\forall \left| \psi \right\rangle \in \mathcal{H}.
\label{eq:Variational_DHS}
\end{equation}
The extremality within the energy spectrum of the sought-after minimal
eigenvalue $E_{0}$ of $\hat{H}$ is absolutely crucial for the effectiveness
and predictive power of variational ground state approaches. It namely
implies that the accuracy of a trial state $\left| \psi \right\rangle $
increases as the variational energy $\left\langle \psi \right| \hat{H}\left|
\psi \right\rangle $ approaches the ground state energy $E_{0}$, and
equality with the exact ground state is reached if and only if the
variational energy equals $E_{0}$''.

From the formulation of the problem we understand that $E_{k}$ and $\left|
\psi _{k}\right\rangle $ are eigenvalues and eigenvectors of $\hat{H}$ and
that we know the $D$-dimensional Hilbert space $\mathcal{H}$ because we are
able to pick up a trial state $\left| \psi \right\rangle $ from it. In such
a case we can also choose a complete orthonormal basis set $\mathcal{B}%
=\left\{ \left| i\right\rangle ,\;i=0,1,\ldots ,D-1\right\} $ for $\mathcal{H%
}$ and build the $D\times D$ matrix representation $\mathbf{H}$ of $\hat{H}$
with matrix elements $H_{ij}=\left\langle i\right| \hat{H}\left|
j\right\rangle $. Consequently, the diagonalization of $\mathbf{H}$ gives us
the \textit{exact} eigenvalues and eigenvectors of $\hat{H}$. Unfortunately,
in practice we never face such an advantageous situation.

The RRM was developed to deal with actual problems in which the Hilbert
space is not finite-dimensional\cite{M33,F22}. In such cases the RRM yields
approximate eigenvalues and eigenvectors and their accuracy increases with
the dimension $N$ of the matrix representation of the Hamiltonian operator.
The exact result is expected to be obtained in the limit $N\rightarrow
\infty $. For this reason, we have called the model introduced by Ding et al
unrealistic.

\section{The Rayleigh-Ritz method}

\label{sec:RRM}

The RRM applies to any Hermitian operator $H$ with eigenvalues $E_{k}$ and
eigenvectors $\left| \psi _{k}\right\rangle $
\begin{equation}
H\left| \psi _{k}\right\rangle =E_{k}\left| \psi _{k}\right\rangle
,\;k=0,1,\ldots .  \label{eq:eigen_eq}
\end{equation}
If we have a complete set of non-orthogonal vectors $\left|
u_{i}\right\rangle $, $i=0,1,\ldots $, then the approximate RRM eigenvalues $%
W_{n}$ are roots of the secular determinant\cite{P68,SO96,F24}
\begin{equation}
\left| \mathbf{H}-W\mathbf{S}\right| =0,  \label{eq:sec_det}
\end{equation}
where $\mathbf{H}$ and $\mathbf{S}$ are $N\times N$ matrices with elements $%
H_{ij}=\left\langle u_{i}\right| H\left| u_{j}\right\rangle $ and $%
S_{ij}=\left\langle u_{i}\right. \left| u_{j}\right\rangle $, $%
i,j=0,1,\ldots N-1$, respectively. It is well known that $W_{k}\geq E_{k}$
for all $k=0,1,\ldots ,N-1$\cite{P68,SO96,M33,F22,F24}. It seems to be
interesting to apply the RRM to the unrealistic problem outlined in section~%
\ref{sec:unrealistic}.

\subsection{Projected Hamiltonian}

\label{subsec:proj_H}

In this subsection we modify the notation of section~\ref{sec:unrealistic}
in order to facilitate the discussion. We assume that we have a Hamiltonian
operator $H$ defined on an infinite-dimensional Hilbert space $\mathcal{H}$
an focus our attention on the projection
\begin{equation}
H_{D}=\sum_{k=0}^{D-1}E_{k}\left| \psi _{k}\right\rangle \left\langle \psi
_{k}\right| ,  \label{eq:H_D}
\end{equation}
of $H$ on a subspace $\mathcal{H}_{D}$ of dimension $D$ spanned by the set
of eigenvectors $\mathcal{S}_{D}=\left\{ \left| \psi _{k}\right\rangle
,\;k=0,1,\ldots ,D-1\right\} $. Obviously, $\mathcal{H}_{D}\subset \mathcal{H%
}$.

In order to apply the RRM to $H_{D}$ we need the matrix elements
\begin{equation}
\left( H_{D}\right) _{ij}=\sum_{k=0}^{D-1}E_{k}\left\langle u_{i}\right.
\left| \psi _{k}\right\rangle \left\langle \psi _{k}\right. \left|
u_{j}\right\rangle ,  \label{eq:(H_D)_ij}
\end{equation}
where $\left\{ \left| u_{j}\right\rangle ,\;j=0,1,\ldots \right\} $ spans
the Hilbert space $\mathcal{H}$. We do not assume the vectors $\left|
u_{j}\right\rangle $ to be orthonormal.

Under the conditions given above the results may be somewhat unexpected. For
example, if $\left| \psi \right\rangle \in \mathcal{H}$ and $\left\langle
\psi _{k}\right. \left| \psi \right\rangle =0$, $k=0,1,\ldots
,n-1,n+1,\ldots ,D-1$, then
\begin{equation}
\frac{\left\langle \psi \right| H_{D}\left| \psi \right\rangle }{%
\left\langle \psi \right. \left| \psi \right\rangle }=E_{n}\frac{\left|
\left\langle \psi _{n}\right. \left| \psi \right\rangle \right| ^{2}}{%
\left\langle \psi \right. \left| \psi \right\rangle }.  \label{eq:<H_D>}
\end{equation}
Upon taking into account the Cauchy-Schwartz inequality\cite{A67} $\left|
\left\langle \psi _{n}\right. \left| \psi \right\rangle \right| ^{2}\leq
\left\langle \psi _{n}\right. \left| \psi _{n}\right\rangle \left\langle
\psi \right. \left| \psi \right\rangle =\left\langle \psi \right. \left|
\psi \right\rangle $ we conclude that.
\begin{eqnarray}
\frac{\left\langle \psi \right| H_{D}\left| \psi \right\rangle }{%
\left\langle \psi \right. \left| \psi \right\rangle } &\leq &E_{n}\;\mathrm{%
if}\;E_{n}>0,  \nonumber \\
\frac{\left\langle \psi \right| H_{D}\left| \psi \right\rangle }{%
\left\langle \psi \right. \left| \psi \right\rangle } &\geq &E_{n}\;\mathrm{%
if}\;E_{n}<0.  \label{eq:UL_bounds}
\end{eqnarray}
We appreciate that it is possible to obtain both upper and lower bounds
instead of only upper ones.

\subsection{Simple example}

\label{subsec:example}

A suitable toy model is given by
\begin{equation}
H=-\frac{1}{2}\frac{d^{2}}{dx^{2}},  \label{eq:H_model}
\end{equation}
with the boundary conditions $\psi (0)=\psi (1)=0$. This example was chosen
in recent discussions of the RRM\cite{F22,F24}. The exact eigenvalues and
eigenfunctions are
\begin{equation}
E_{k}=\frac{k^{2}\pi ^{2}}{2},\;\psi _{k}(x)=\sqrt{2}\sin \left( k\pi
x\right) ,\;k=1,2,\ldots .  \label{eq:E_k,psi_k_example}
\end{equation}

For simplicity, we choose the non-orthogonal basis set
\begin{equation}
u_{i}(x)=x^{i}(1-x),\;i=1,2,\ldots ,  \label{eq:u_i(x)}
\end{equation}
already used earlier\cite{F24}. Table~\ref{tab:RRH} shows that the RRM
eigenvalues $W_{k}$ converge from above towards the exact eigenvalues $E_{k}$
as expected\cite{P68,SO96,M33,F22,F24}.

Tables \ref{tab:D=1} and \ref{tab:D=2} show the RRM results for the
projected Hamiltonian $H_{D}$ with $D=1$ and $D=2$, respectively (note that $%
k=1,2,\ldots ,D$ in this example). We appreciate that there are $D$
eigenvalues $W_{k}\neq 0$, $k=1,2,\ldots ,D$, and the remaining roots vanish
$W_{k}=0$, $D<k\leq N$ as expected. It is interesting that the RRM yields
the eigenvalues $E=0$ exactly while the others are approximate (though, they
converge towards the exact ones as $N$ increases). The reason is that the
RRM yields $N-D$ approximate solutions $\left| \varphi _{j}\right\rangle $
that are orthogonal to $\left| \psi _{k}\right\rangle $, $k=1,2,\ldots ,D$.
We appreciate that in these particular cases the RRM eigenvalues $W_{k}$
converge towards the exact ones $E_{k}$ from below. However, this result is
not general. In table~\ref{tab:D=3} we show RRM eigenvalues for $D=3$. We
see that this approach yields an upper bound to $E_{1}$ for $N=1$ and lower
bounds to all the eigenvalues for $N>1$. The well known proofs for the upper
bounds mentioned above \cite{M33,F22} (and references therein) do not apply
here because the functions $u_{i}(x)$ cannot be expressed in terms of the
finite set $\left\{ \psi _{k}(x),\;k=1,2,\ldots ,D\right\} $.

For every RRM eigenvalue $W_{k}$ we obtain an approximate solution\cite
{P68,SO96,F24}
\begin{equation}
\left| \varphi _{k}\right\rangle =\sum_{j=1}^{N}c_{jk}\left|
u_{j}\right\rangle ,  \label{eq:varphi_k}
\end{equation}
and we commonly choose such solutions to be orthonormal $\left\langle
\varphi _{k}\right. \left| \varphi _{n}\right\rangle =\delta _{kn}$. From
the results of tables \ref{tab:D=1}, \ref{tab:D=2} and \ref{tab:D=3} and $%
\left\langle \varphi _{i}\right| H_{D}\left| \varphi _{j}\right\rangle
=W_{i}\delta _{ij}$\cite{P68,SO96,F24} we conclude that
\begin{equation}
\sum_{i=1}^{N}\sum_{j=1}^{N}\left\langle \varphi _{i}\right| H_{D}\left|
\varphi _{j}\right\rangle \left| \varphi _{i}\right\rangle \left\langle
\varphi _{j}\right| =\sum_{i=1}^{D}W_{i}\left| \varphi _{i}\right\rangle
\left\langle \varphi _{i}\right| .  \label{eq:spectral_H_D}
\end{equation}

In order to investigate the particular case $N=1$ in more detail we
calculated the expectation value $\left\langle H_{D}\right\rangle $ with the
function $u_{1}(x)$. The results in table~\ref{tab:N=1} clearly show that $%
\left\langle H_{1}\right\rangle <E_{1}<\left\langle H\right\rangle =5$ and $%
E_{1}<\left\langle H_{D}\right\rangle <\left\langle H\right\rangle $ for $%
D>1 $. The table suggests the obvious conclusion that $\lim\limits_{D%
\rightarrow \infty }\left\langle H_{D}\right\rangle =\left\langle
H\right\rangle $ which can be proved analytically:
\begin{equation}
\sum_{k=1}^{\infty }\frac{k^{2}\pi ^{2}}{2}\frac{\left\langle u_{1}\right.
\left| \psi _{k}\right\rangle \left\langle \psi _{k}\right. \left|
u_{1}\right\rangle }{\left\langle u_{1}\right. \left| u_{1}\right\rangle }=%
\frac{240}{\pi ^{4}}\sum_{k=1}^{\infty }\frac{1-(-1)^{k}}{k^{4}}=5.
\end{equation}

It is worth adding that for $D=4$ we have $W_{1}>E_{1}$ for $N=1$, as
already pointed out for $D=3$, and also $W_{2}>E_{2}$ for $N=3$. The RRM
eigenvalues converge from below for all $N>3$.

We also carried out some numerical experiments with $E_{k}=k^{2}\pi ^{2}/2+c$%
, where $c$ is a real constant. Table~\ref{tab:D=3_shifted} shows that for $%
c=-5$, $W_{1}$ becomes an upper bound while $W_{2}$ and $W_{3}$ remain lower
bounds for all $N$.

\subsection{Arbitrary example}

\label{subsec:arbitrary}

We can extend the results of the preceding subsection to a more general case
based on the same toy model. Suppose that we have a set of orthonormal
vectors $\left| \psi _{k}\right\rangle $, $k=$ $1,2,\ldots ,D$ and construct
the projected Hamiltonian operator
\begin{equation}
H_{D}=\sum_{k=1}^{D}\alpha _{k}\left| \psi _{k}\right\rangle \left\langle
\psi _{k}\right| ,  \label{eq:H_D_general}
\end{equation}
where $\alpha _{k}$ are arbitrary real numbers. We can obviously apply the
RRM as in the preceding example. Table~\ref{tab:H_D_gen} shows RRM
eigenvalues for the case $D=3$ and $\alpha _{k}=k$. In this case we
appreciate that $W_{1}>1$ for $N=1$ and $W_{1}<1$ for all $N>1$. Also $%
W_{2}>2$ for $N=3$ and $W_{2}<2$ for all $N>3$. We conclude that the RRM
yields upper and lower bounds for this kind of projected Hamiltonian
operators.

\subsection{What about the identity operator?}

\label{subsec:identity_proj}

In the calculations discussed above we have considered the identity operator
on $\mathcal{H}$
\begin{equation}
I=\sum_{k=1}^{\infty }\left| \psi _{k}\right\rangle \left\langle \psi
_{k}\right| ,  \label{eq:I}
\end{equation}
so that the overlap matrix is given by $S_{ij}=\left\langle u_{i}\right|
I\left| u_{j}\right\rangle =\left\langle u_{i}\right. \left|
u_{j}\right\rangle $, $i,j=1,2,\ldots ,N$. We obtain completely different
results if we consider the projection of $I$ on $\mathcal{H}_{D}$
\begin{equation}
I_{D}=\sum_{k=1}^{D}\left| \psi _{k}\right\rangle \left\langle \psi
_{k}\right| ,  \label{eq:I_D}
\end{equation}
and the overlap matrix $S_{i,j}^{D}=\left\langle u_{i}\right| I_{D}\left|
u_{j}\right\rangle $, $i,j=1,2,\ldots ,N$. In this case, we obtain upper
bounds for $1\leq N<D$ and the exact eigenvalues when $N=D$. For example,
for $D=N=5$ we have
\begin{equation}
\left| \mathbf{H}_{5}-W\mathbf{S}^{5}\right| =\frac{67108864\left( \pi
^{2}-2W\right) \left( 2\pi ^{2}-W\right) \left( 8\pi ^{2}-W\right) \left(
9\pi ^{2}-2W\right) \left( 25\pi ^{2}-2W\right) }{9765625\pi ^{46}},
\label{eq:sec_det_N=D=5}
\end{equation}
that yields the fifth lowest eigenvalues exactly. In this case $\left|
\mathbf{H}_{D}-W\mathbf{S}^{D}\right| =\left| \mathbf{H}_{D}\right| =\left|
\mathbf{S}^{D}\right| =0$ for all $N>D$.

\section{Conclusions}

\label{sec:conclusions}

It is not clear to us why Ding et al\cite{DHS24} introduced the variational
principle by means of a projected Hamiltonian operator because this
principle as well as the properties of the RRM have already been proved for
the kind of operators commonly found in actual physical problems\cite
{M33,F22}. However, when the RRM is applied to the projected Hamiltonian in
terms of a basis set for $\mathcal{H}\supset \mathcal{H}_{D}$, one obtains
lower bounds in most cases as shown above. It is interesting that the null $%
N-D$ eigenvalues are given exactly for all $N>D$ although the approximate
RRM solutions $\left| \varphi _{j}\right\rangle $ only yield the exact ones $%
\left| \psi _{k}\right\rangle $ in the limit $N\rightarrow \infty $. As
argued above, it is only necessary that the approximate RRM solutions are
orthogonal to the exact ones $\left| \psi _{k}\right\rangle $ in $\mathcal{H}%
_{D}$. We have also shown that the RRM yields upper bounds for $1\leq N<D$
and exact eigenvalues for $N=D$ when we consider the identity operator (\ref
{eq:I_D}) instead of (\ref{eq:I}).

\begin{table}[tbp]
\caption{Lowest RRM eigenvalues for the Hamiltonian (\ref{eq:H_model})}
\label{tab:RRH}
\begin{center}
\par
\begin{tabular}{D{.}{.}{3}D{.}{.}{11}D{.}{.}{11}D{.}{.}{11}D{.}{.}{11}}
\hline \multicolumn{1}{l}{$N$}&\multicolumn{1}{c}{$W_1$}&
\multicolumn{1}{c}{$W_2$} &
\multicolumn{1}{c}{$W_3$} & \multicolumn{1}{c}{$W_4$} \\
\hline
  1&   5           &              &              &               \\
  3&   4.934874810 &  21          &  51.06512518 &               \\
  5&   4.934802217 &  19.75077640 &  44.58681182 &  100.2492235  \\
  7&   4.934802200 &  19.73923669 &  44.41473408 &  79.99595777  \\
  9&   4.934802200 &  19.73920882 &  44.41322468 &  78.97848206  \\
 11&   4.934802200 &  19.73920880 &  44.41321981 &  78.95700917  \\
 13&   4.934802200 &  19.73920880 &  44.41321980 &  78.95683586  \\
 15&   4.934802200 &  19.73920880 &  44.41321980 &  78.95683521  \\
 17&   4.934802200 &  19.73920880 &  44.41321980 &  78.95683520  \\
 19&   4.934802200 &  19.73920880 &  44.41321980 &  78.95683520  \\

 \end{tabular}
\par
\end{center}
\end{table}

\begin{table}[tbp]
\caption{RRM for the projected Hamiltonian with $D=1$}
\label{tab:D=1}
\begin{center}
\par
\begin{tabular}{D{.}{.}{3}D{.}{.}{11}D{.}{.}{4}}
\hline \multicolumn{1}{l}{$N$}&
\multicolumn{1}{c}{$W_1$}&\multicolumn{1}{c}{$W_k,\ 1< k\leq N$} \\
\hline
  1  &  4.927671482  &     \\
  3  &  4.934799721  &  0   \\
  5  &  4.934802200  &  0   \\
  7  &  4.934802200  &  0   \\

 \end{tabular}
\par
\end{center}
\end{table}

\begin{table}[tbp]
\caption{RRM for the projected Hamiltonian with $D=2$}
\label{tab:D=2}
\begin{center}
\par
\begin{tabular}{D{.}{.}{3}D{.}{.}{11}D{.}{.}{11}D{.}{.}{4}}
\hline \multicolumn{1}{l}{$N$}&
\multicolumn{1}{c}{$W_1$} &
\multicolumn{1}{c}{$W_2$}&\multicolumn{1}{c}{$W_k,\ 2< k\leq N$} \\
\hline
  1 &  4.927671482 &    &    \\
  3 &  4.934799721 &  19.40270646  &  0  \\
  5 &  4.934802200 &  19.73799899  &  0  \\
  7 &  4.934802200 &  19.73920734  &  0  \\
  9 &  4.934802200 &  19.73920880  &  0  \\
 11 &  4.934802200 &  19.73920880  &  0  \\
 \end{tabular}
\par
\end{center}
\end{table}

\begin{table}[tbp]
\caption{RRM for the projected Hamiltonian with $D=3$}
\label{tab:D=3}
\begin{center}
\par
\begin{tabular}{D{.}{.}{3}D{.}{.}{11}D{.}{.}{11}D{.}{.}{11}D{.}{.}{4}}
\hline \multicolumn{1}{l}{$N$}&
\multicolumn{1}{c}{$W_1$} &
\multicolumn{1}{c}{$W_2$} & \multicolumn{1}{c}{$W_3$}&\multicolumn{1}{c}{$W_k,\ 3< k\leq N$} \\
\hline
  1  &  4.988506932 &              &              &    \\
  3  &  4.934799541 &  19.40270646 &  41.72191568 &  0 \\
  5  &  4.934802200 &  19.73799899 &  44.37877225 &  0 \\
  7  &  4.934802200 &  19.73920734 &  44.41306667 &  0 \\
  9  &  4.934802200 &  19.73920880 &  44.41321950 &  0 \\
 11  &  4.934802200 &  19.73920880 &  44.41321980 &  0 \\
 13  &  4.934802200 &  19.73920880 &  44.41321980 &  0 \\

 \end{tabular}
\par
\end{center}
\end{table}

\begin{table}[tbp]
\caption{Expectation value of $H_D$ with $u_1(x)$}
\label{tab:N=1}
\begin{center}
\par
\begin{tabular}{D{.}{.}{3}D{.}{.}{11}}
\hline \multicolumn{1}{l}{$D$}& \multicolumn{1}{c}{$\langle H_D
\rangle$}
\\
\hline
  1&  4.927671482 \\
  3&  4.988506932 \\
  5&  4.996391207 \\
  7&  4.998443548 \\
  9&  4.999194603 \\
 11&  4.999531169 \\
 13&  4.999703701 \\
 15&  4.999801038 \\
 17&  4.999860037 \\
 19&  4.999897849 \\
 21&  4.999923186 \\
 23&  4.999940795 \\
 25&  4.999953410 \\
 27&  4.999962682 \\
 29&  4.999969649 \\
 31&  4.999974985 \\
 33&  4.999979140 \\
 35&  4.999982424 \\
 37&  4.999985053 \\
 39&  4.999987183 \\
 41&  4.999988927 \\
 43&  4.999990368 \\
 45&  4.999991570 \\

 \end{tabular}
\par
\end{center}
\end{table}

\begin{table}[tbp]
\caption{RRM for the projected Hamiltonian with $D=3$ and $E_k=k^2\pi^2/2 -
5 $}
\label{tab:D=3_shifted}
\begin{center}
\par
\begin{tabular}{D{.}{.}{3}D{.}{.}{11}D{.}{.}{11}D{.}{.}{11}D{.}{.}{4}}
\hline \multicolumn{1}{l}{$N$}& \multicolumn{1}{c}{$W_1$} &
\multicolumn{1}{c}{$W_2$} & \multicolumn{1}{c}{$W_3$}&\multicolumn{1}{c}{$W_k,\ 3< k\leq N$} \\
\hline
  1  &  -0.01111693899 &              &              &    \\
  3  &  -0.06519776461 &  14.48794350 &  37.02490009 &  0 \\
  5  &  -0.06519779945 &  14.73830544 &  39.38265032 &  0 \\
  7  &  -0.06519779945 &  14.73920771 &  39.41308391 &  0 \\
  9  &  -0.06519779945 &  14.73920880 &  39.41321953 &  0 \\
 11  &  -0.06519779945 &  14.73920880 &  39.41321980 &  0 \\
 13  &  -0.06519779945 &  14.73920880 &  39.41321980 &  0 \\

 \end{tabular}
\par
\end{center}
\end{table}

\begin{table}[tbp]
\caption{RRM for the arbitrary Hamiltonian (\ref{eq:H_D_general}) with $D=3$
and $\alpha_k=k$}
\label{tab:H_D_gen}
\begin{center}
\par
\begin{tabular}{D{.}{.}{3}D{.}{.}{11}D{.}{.}{11}D{.}{.}{11}D{.}{.}{4}}
\hline \multicolumn{1}{l}{$N$}& \multicolumn{1}{c}{$W_1$} &
\multicolumn{1}{c}{$W_2$} & \multicolumn{1}{c}{$W_3$}&\multicolumn{1}{c}{$W_k,\ 3< k\leq N$} \\
\hline
  1  &  1.002664294  &              &              &   \\
  3  &  0.9999994479 &  2.027339721 &  2.818209291 &   \\
  5  &  0.9999999999 &  1.999877421 &  2.997673155 &  0 \\
  7  &  0.9999999999 &  1.999999852 &  2.999989656 &  0 \\
  9  &  0.9999999999 &  1.999999999 &  2.999999979 &  0 \\
 11  &  0.9999999999 &  1.999999999 &  2.999999999 &  0 \\
 13  &  1.0000000000 &  1.999999999 &  2.999999999 &  0 \\

 \end{tabular}
\par
\end{center}
\end{table}

\end{document}